\documentclass[prb,preprint]{revtex4-1}

\usepackage{amsmath} 
\usepackage{graphicx} 
\usepackage{subfigure} 
\usepackage{endnotes}

\begin{document}

\title{Gravitational-wave science in the high school classroom}
\author{Benjamin Farr\endnote{Electronic mail:bfarr@u.northwestern.edu}}
  \affiliation{Northwestern University, Department of Physics \& Astronomy, Evanston, IL 60208}
\author{GionMatthias Schelbert\endnote{Electronic mail:schelbertg@eths.k12.il.us}}
  \affiliation{Evanston Township High School, Science Department, Evanston, IL 60202}
\author{Laura Trouille\endnote{Electronic mail:l-trouille@northwestern.edu}}
  \affiliation{Northwestern University, Center for Interdisciplinary Exploration and Research in Astrophysics, Evanston, IL 60208}

\begin{abstract}
This article describes a set of curriculum modifications designed to integrate gravitational-wave science into a high school physics or astronomy curriculum. Gravitational-wave scientists are on the verge of being able to detect extreme cosmic events, like the merger of two black holes, happening hundreds of millions of light years away. Their work has the potential to propel astronomy into a new era by providing an entirely new means of observing astronomical phenomena. Gravitational-wave science encompasses astrophysics, physics, engineering, and quantum optics. As a result, this curriculum exposes students to the interdisciplinary nature of science. It also provides an authentic context for students to learn about astrophysical sources, data analysis techniques, cutting-edge detector technology, and error analysis.
\end{abstract}

\maketitle

\section{Introduction}
Today the vast field of astronomy exists almost entirely within one medium of observation: electromagnetic radiation.  From radio waves to gamma rays, the electromagnetic spectrum has provided us with the observational data necessary to reach our current understanding of the universe.  However, this restricted view of the universe has provided us with a relatively limited knowledge of objects that emit little to no light, such as black holes and neutron stars.  To better observe these bodies, we must look to an alternative cosmic messenger.  The Laser Interferometer Gravitational-wave Observatory (LIGO)\cite{LIGOstatus} and partner observatory Virgo\cite{VIRGOstatus} are on the verge of making the first direct detection of gravitational waves, which will provide astronomers with the most direct observations of black holes yet.

This cutting-edge field was brought into the high school classroom as part of the Graduate STEM Fellows in K-12 Education (GK-12) program, funded by the National Science Foundation.  Authors B. Farr (GK-12 fellow and LIGO Scientific Collaboration member) and G. Shelbert (GK-12 partner and astronomy teacher) spent the 2010--2011 academic year developing and integrating lessons based on gravitational-wave science into the high school astronomy curriculum, under the guidance and mentorship of L. Trouille (GK-12 mentor).  The course had no prerequisites, and students' math proficiency ranged from basic algebra to multivariate calculus.  However, most of the students had never heard of gravitational-wave science and therefore started with the same prior knowledge of the subject.  Students were fascinated by the concept of gravitational waves and colliding black holes, which resulted in significant self-motivation.  By explicitly connecting existing topics in the curriculum to gravitational-wave science whenever possible, these lessons gave coherency to units that before may have seemed disconnected.

The structure of this paper is as follows.  In Section~\ref{sec:GWscience} we provide an introduction to gravitational waves, detectors, sources, and data analysis techniques.  Section~\ref{sec:classroom} outlines several examples of how we have incorporated gravitational-wave science into the high school astronomy classroom.

\section{Gravitational-wave Science}
  \label{sec:GWscience}
Gravitational-wave science is the study of small ripples in space and time emitted by the acceleration of massive bodies. In the following sections we provide background information on the physics underlying the emission of gravitational waves (Section~\ref{subsec:GWaves}), the main detection methods (Section~\ref{subsec:detectors}), the astronomical phenomena our gravitational-wave detectors are sensitive to (Section~\ref{subsec:sources}), and the main methods for data analysis (Section~\ref{subsec:DA}).

\subsection{Gravitational waves}
  \label{subsec:GWaves}
\begin{figure}[ht]
  \subfigure[~$t=P/4$]{
   \includegraphics[scale=0.3]{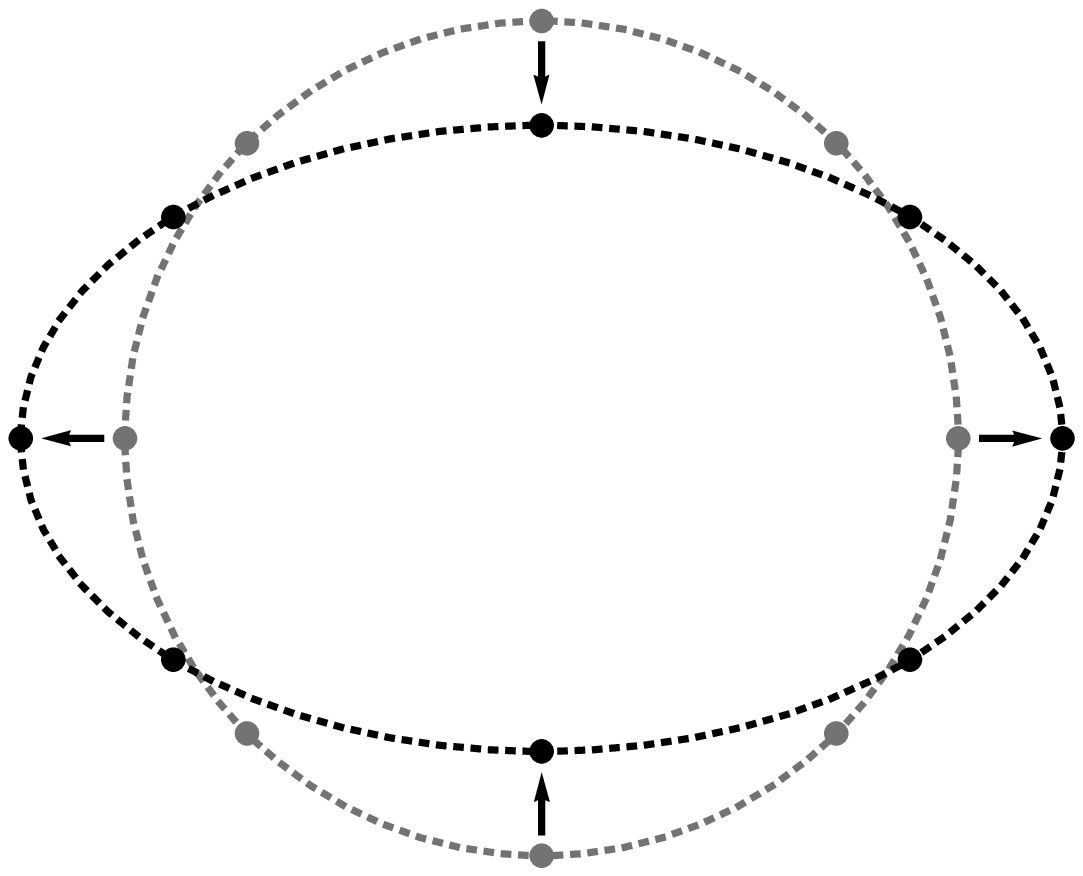}
    \label{subfig:ring1}
  }
  \subfigure[~$t=P/2$]{
   \includegraphics[scale=0.3]{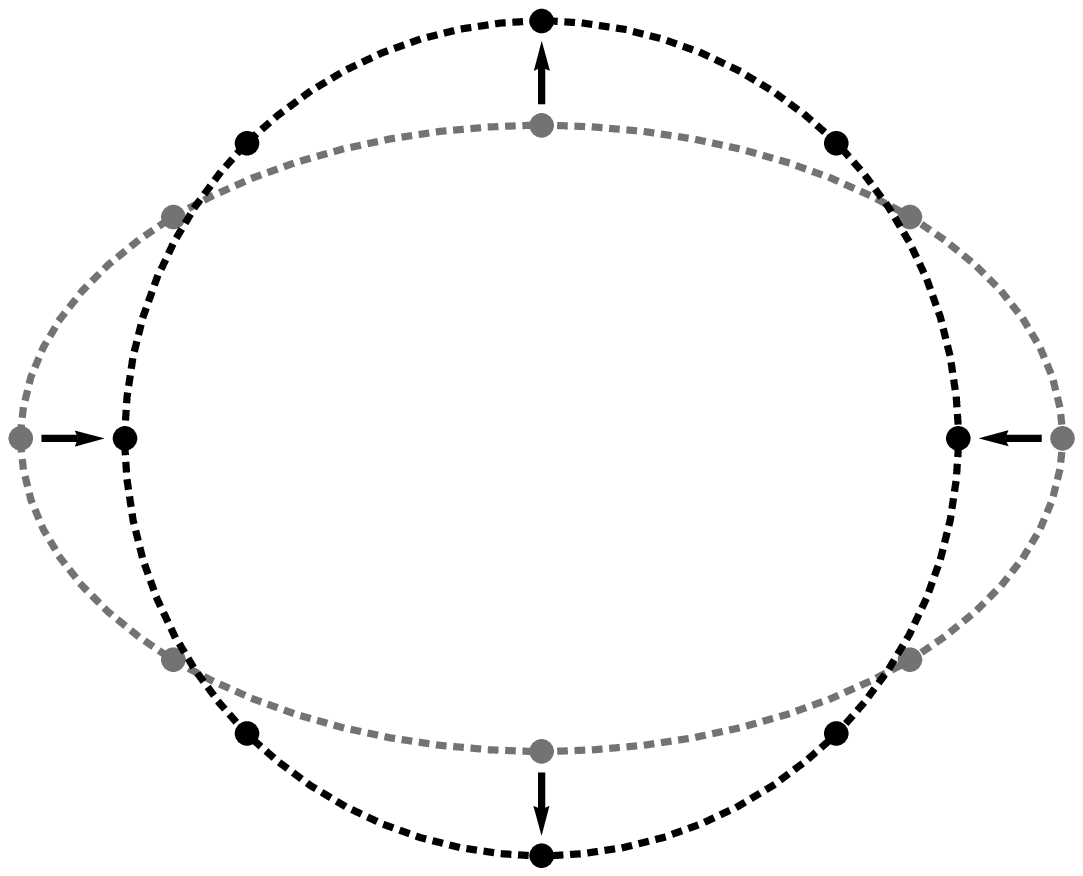}
    \label{subfig:ring2}
  }
  \subfigure[~$t=3P/4$]{
   \includegraphics[scale=0.3]{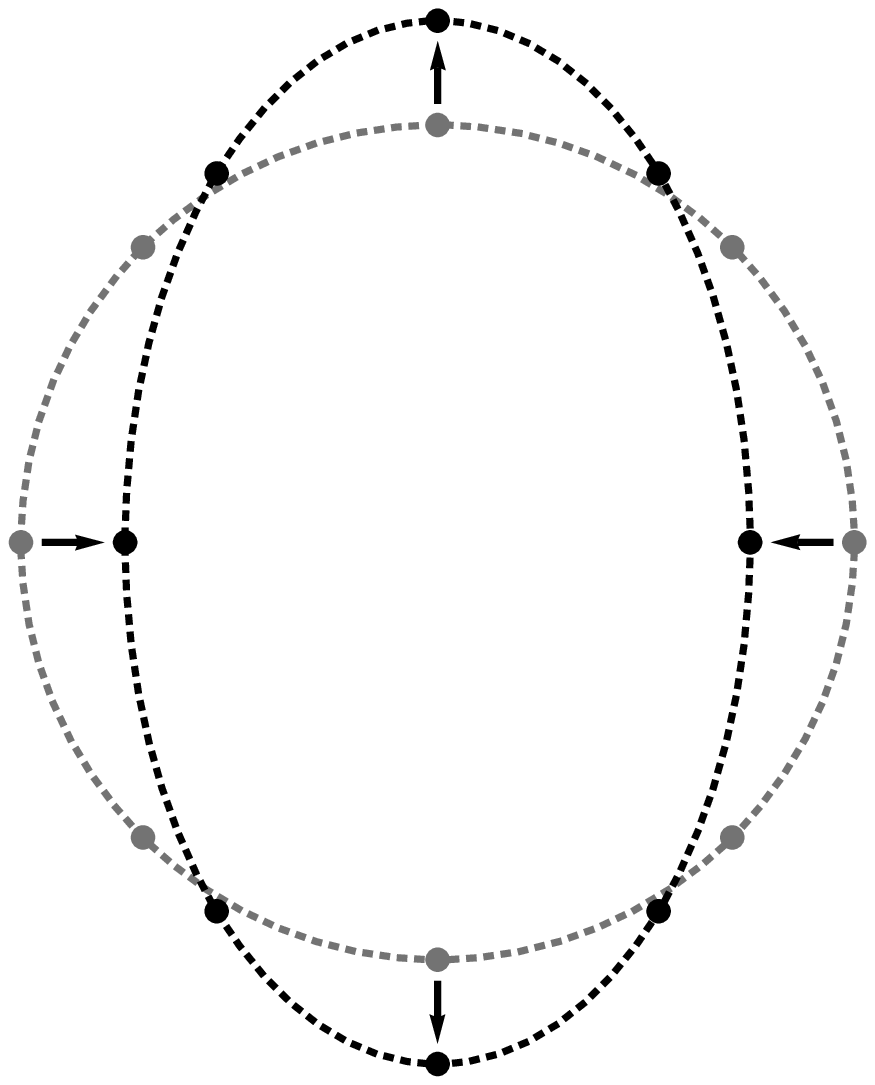}
    \label{subfig:ring3}
  }
  \subfigure[~$t=P$]{
   \includegraphics[scale=0.3]{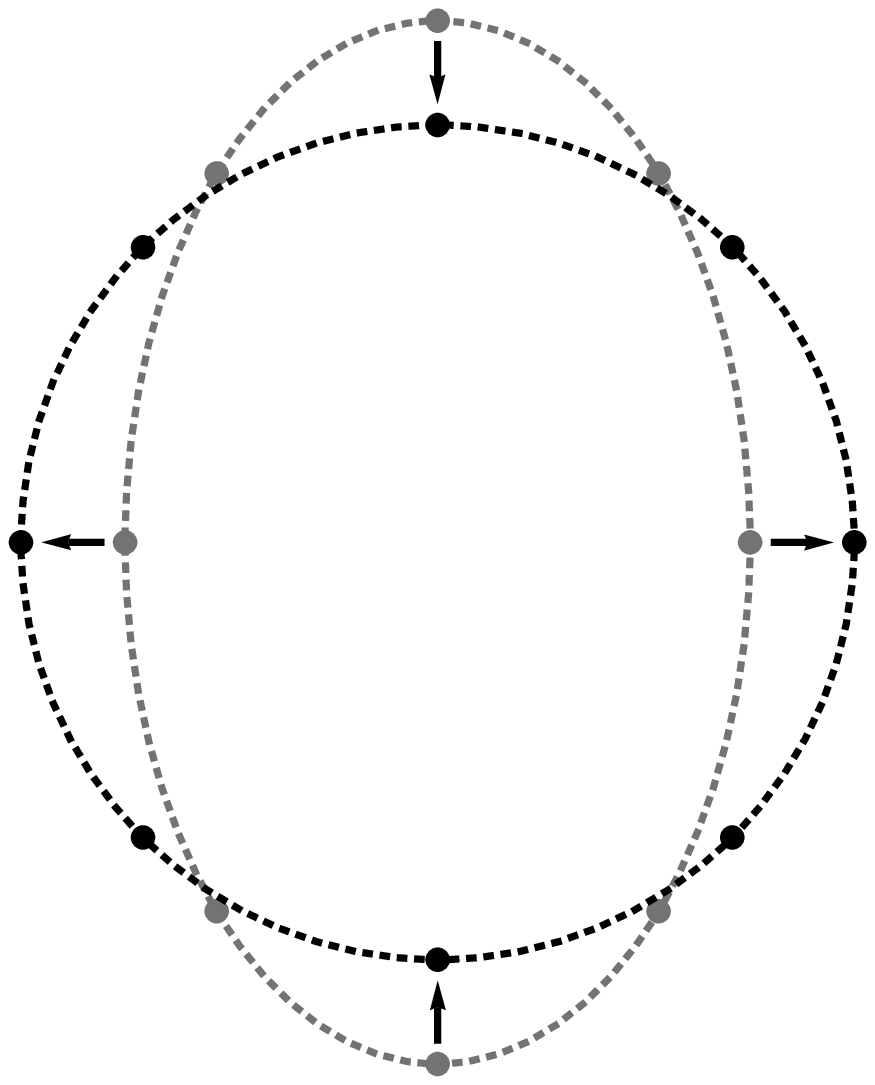}
    \label{subfig:ring4}
  }
  \caption{Time evolution of a ring of particles influenced by a passing gravitational wave, where $P$ is the period of the wave.  Each consecutive panel is a snapshot taken $P/4$ later in time.  In each figure, both the current state of the ring (black) and the previous state of the ring (gray) are shown.}
\end{figure}

With his general theory of relativity,\cite{Relativity} Einstein triggered the most significant advancement in our understanding of gravity since Newton.  Einstein's theory proposes that the dimension of time can be treated much like our three spatial dimensions, which together constitute spacetime.  This spacetime is influenced by the presence of mass in a way similar to a stretched fabric holding a heavy object.  When other massive bodies travel through this curved region of space, their motions deviate from the normally straight paths, like a ball rolling on a curved fabric.  This analogy can be extended further by considering the rapid movement of very massive objects on the fabric, which produces ripples traveling outward from the bodies, as also happens in spacetime.\cite{Hartle}  These ripples produced by the acceleration of massive objects traveling through spacetime are gravitational waves, and they carry with them a wealth of information about their source.  As these gravitational waves propagate, they exert a periodic expansion and contraction of the spacetime they pass through in directions perpendicular to the direction of travel.  To examine the effects of these waves locally, consider a ring of particles floating in space, free of any external forces.  As the wave passes, the distance along each axis undergoes periodic expansion and contraction in a fashion exactly opposite to that of the perpendicular axis (see Fig.~\ref{subfig:ring1}--\subref{subfig:ring4}).  The goal of gravitational-wave detectors is to measure these minuscule vibrations, with the hope of learning more about their sources.

\subsection{Gravitational-wave detectors}
  \label{subsec:detectors}

\begin{figure}[ht]
 \includegraphics[scale=0.5]{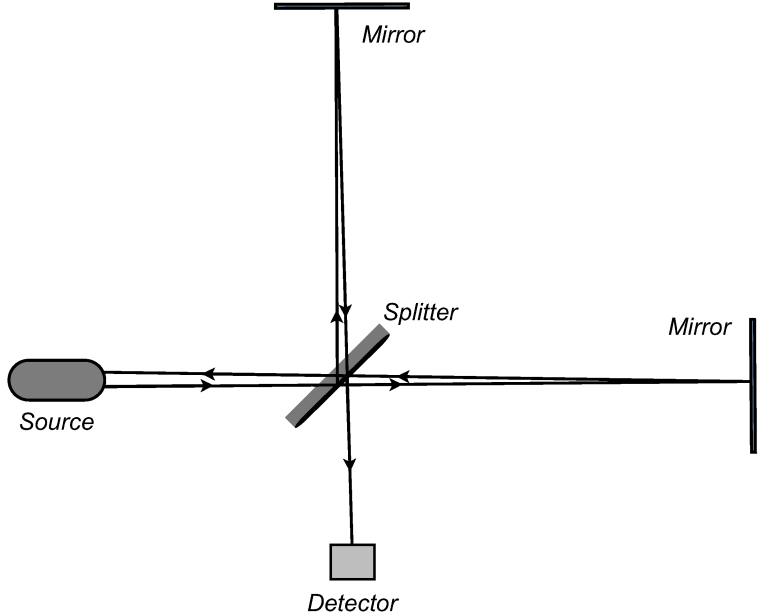}
  \caption{\label{fig:michelson}Schematic diagram of the Michelson interferometer, showing the path of the light beam as it is split and then recombined before entering the photodetector. The dark line on the beam splitter indicates the reflective coating, and beams reflected from different sides of the coating have a $180^\circ$ phase difference.  This setup is very similar to that used in gravitational-wave detector technology.}
\end{figure}

Currently the most sensitive operational gravitational-wave detectors are based on the Michelson interferometer, which uses the interference properties of light to make incredibly precise measurements of distances.\cite{LIGO}  As shown in Fig.~\ref{fig:michelson}, these detectors split a coherent light beam from a single laser into two beams. These two beams travel along different paths before recombining and entering the photodetector. More specifically, the detector is set up in an `L' formation, with a mirror suspended at the end of each arm. A laser emits a beam of light that is \emph{in phase}, meaning the peaks and troughs of each light wave are aligned. This original beam of light is split by the beamsplitter. One beam is reflected off of one mirror while the other beam is reflected off of the other mirror. Once the two beams return to the beamsplitter they recombine, with some light going back toward the laser while the rest passes to the photodetector. The beam splitter has a reflective coating on one side of it, which means that of the two possible paths the light could take to the photodetector, one has reflected from the glass side of the coating, and the other from the vacuum side.  These two different cases of reflection will result in a $180^\circ$ phase difference. If the distances traveled by each half of the beam are equal (i.e., the arms are equal in length), the two light beams will be exactly out of phase and cancel each other out, resulting in all light traveling back toward the laser and none reaching the photodetector. If instead one arm is slightly shorter than the other, the light no longer exactly cancels out, and a nonzero light intensity is measured by the photodetector. The intensity of this light measured at the detector depends very sensitively on the phase difference of the two halves of the beam.  Thus by monitoring the fluctuations in the intensity of exiting light, the difference in arm lengths can be determined with incredible accuracy.

\begin{figure}[ht]
\subfigure[~LIGO Hanford]{
\includegraphics[scale=1.0]{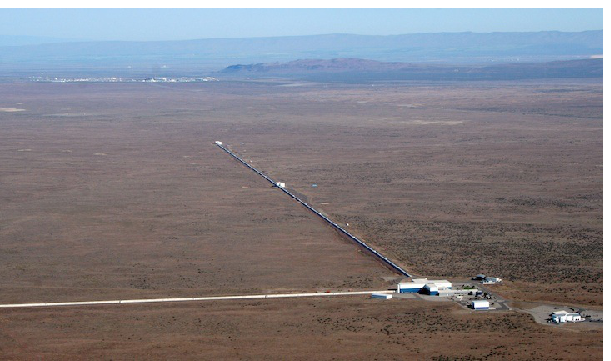}
\label{subfig:LHO}
}
\subfigure[~LIGO Livingston]{
\includegraphics[scale=1.0]{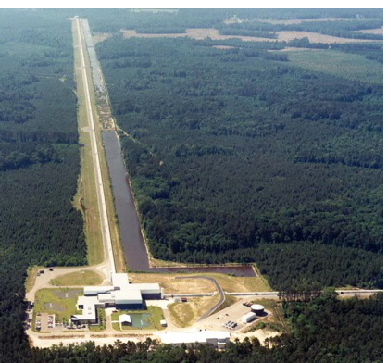}
\label{subfig:LLO}
}
\subfigure[~Virgo]{
\includegraphics[scale=0.88]{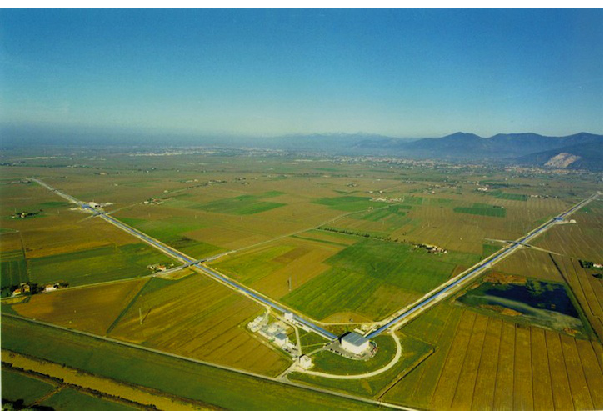}
\label{subfig:Virgo}
}
\caption{\label{fig:IFOs}The LIGO-Virgo gravitational-wave detector network, consisting of \subref{subfig:LHO} LIGO Hanford (Credit: LIGO Laboratory), \subref{subfig:LLO} LIGO Livingston (Credit: LIGO Laboratory), and \subref{subfig:Virgo} Virgo (Credit: EGO).  LIGO Hanford (Hanford, WA) and LIGO Livingston (Livingston, LA) both have 4 km long arms, while Virgo (Cascina, Italy) has 3 km long arms.}
\end{figure}

This is the underlying principle for the detection of gravitational waves.  The three largest detectors in the world based on this design make up the LIGO-Virgo Collaboration (LVC).  Figure~\ref{fig:IFOs} shows the LVC network, consisting of two detectors in the United States (located in Hanford, WA and Livingston, LA) with arm lengths of 4 km that make up the Laser Interferometer Gravitational-wave Observatory (LIGO),\cite{LIGOstatus} as well as Virgo,\cite{VIRGOstatus} a 3 km detector in Italy.

\subsection{Sources}
  \label{subsec:sources}
According to the theory of general relativity, any mass that is accelerating in a way that is not perfectly spherically or cylindrically symmetric will produce gravitational waves.  Though this excludes some processes like the spherically symmetric pulsations of stars, it does include countless other events, ranging from the energetic collision of stars and black holes to the less spectacular toss of a ball.

\begin{figure}[ht]
    \subfigure[~Accurate Model]{
   \includegraphics[scale=0.25]{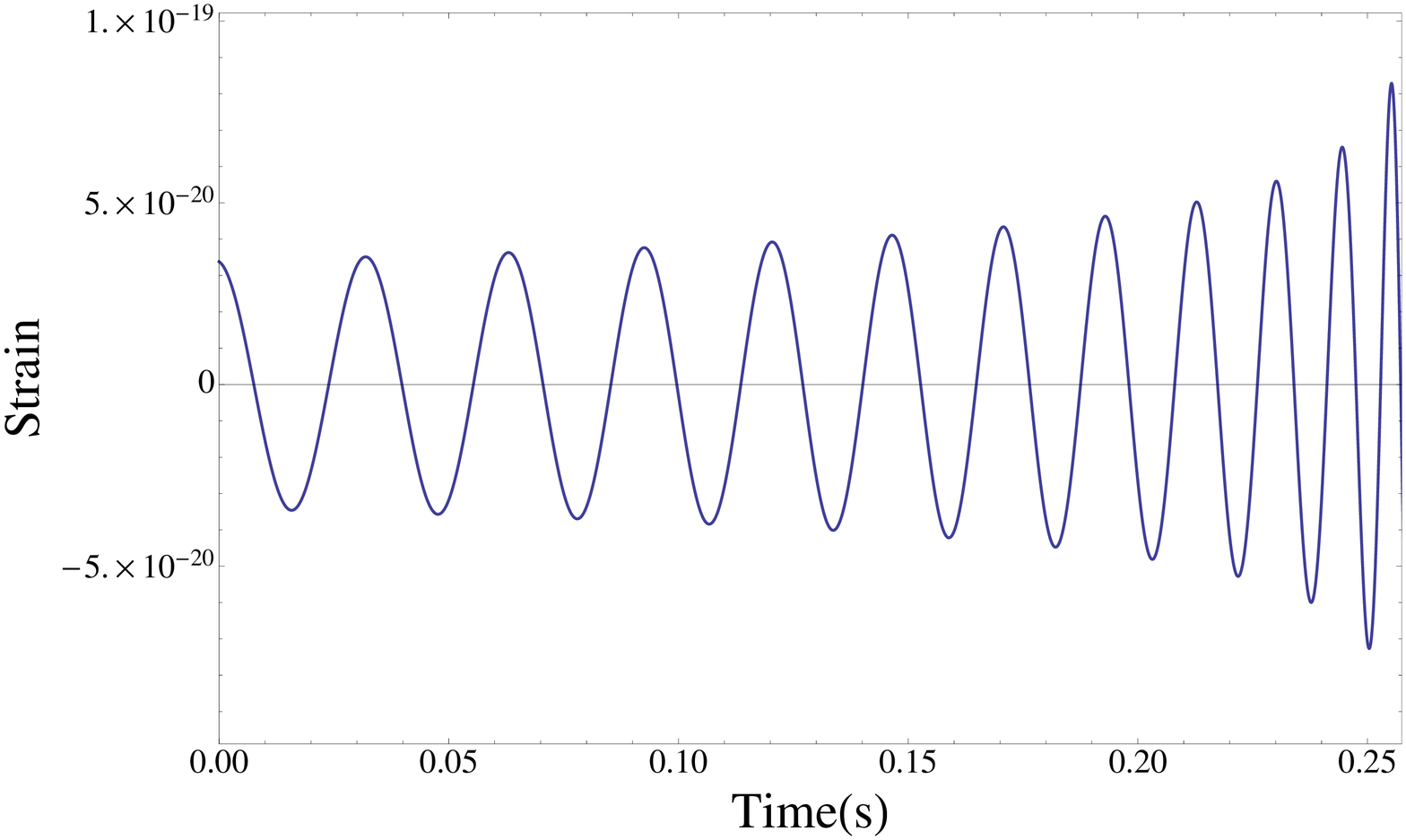}
    \label{subfig:accurate}
    }
    \subfigure[~Common Misconception]{
   \includegraphics[scale=0.25]{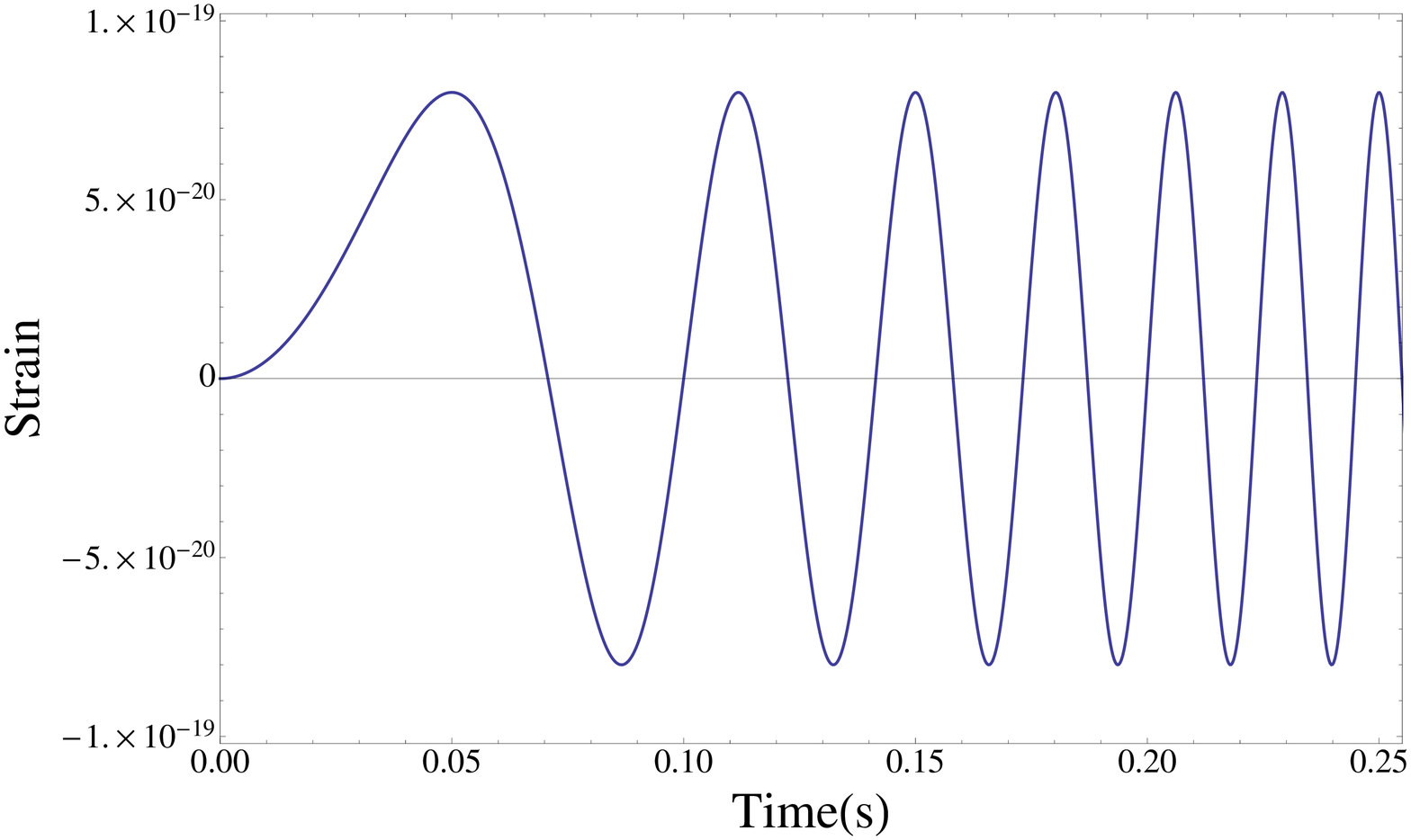}
    \label{subfig:misconception}
    }
    \caption{\label{fig:waveforms}The two most common responses by students when asked to hypothesize what the gravitational wave from a binary merger should look like. \subref{subfig:accurate} A qualitatively accurate model, similar to the majority of student responses.  \subref{subfig:misconception} A model with constant amplitude, the most common misconception among students.}
\end{figure}

Consider a system of two bodies, each about as massive as the Sun, orbiting about one another.  As the bodies orbit, gravitational waves are emitted with a period that is proportional to that of the orbit .  These waves carry energy away from the system.  As the orbit loses energy, the separation between the two objects must shrink, thereby decreasing the orbital period.  Furthermore, as the bodies get closer together, the second time derivative of the quadrupole moment varies more rapidly, resulting in an increase in the gravitational-wave amplitude. This increase in frequency and amplitude continues until the orbital radius decreases to the point of merger, where the two objects physically combine to form a single body.  Until the time of merger the system is said to be in its \emph{inspiral phase}, which is modeled fairly accurately by making small corrections to the non-general relativistic equations of motion.  An example of a gravitational wave produced during the inspiral phase of such a system is shown in Fig.~\ref{subfig:accurate}.  As discussed in Section~\ref{subsec:GR}, when presenting this in a high school classroom setting, the explanation for why the amplitude increases is more qualitative.  The focus is on the students appreciating the connection between the more extreme curvature of spacetime as the bodies get closer together and the increase in amplitude of the signal.

If we were to take the same system, but compress each object's mass into a smaller radius, the inspiral phase would be prolonged.  In this case the orbital radius is able to reach even smaller values before these denser objects merge, thereby increasing the amplitude and frequency reached by the gravitational wave before merger.  Consequently only binary systems containing the densest objects in the universe, namely neutron stars and black holes, are capable of producing gravitational waves at amplitudes and frequencies detectable by current detectors.

The fact that we have yet to detect a gravitational wave, despite being surrounded by sources, is due primarily to the ``stiffness'' of spacetime.  The stiffness of spacetime refers to the incredible amounts of energy in gravitational waves that are required to distort spacetime to a degree we can measure with our detectors. A second major factor is the relatively low amount of energy emitted in gravitational waves by systems in the first place.  As an example of the latter, the amount of energy per second (power) radiated in gravitational waves by the orbit of Jupiter around the Sun is 5200 watts.\cite{jupiter} Even though this is the most energetic source of gravitational waves in our solar system, the energy radiated in all directions each year by the orbit of Jupiter would only be enough to power a single typical household in the United States,\cite{energy}$^,$\endnote{This calculation uses the average energy consumed per household during 2009, averaged over all households in the United States, and excludes biomass energy uses (e.g. wood, coal, solar).} meaning we must look for much more energetic events occuring outside of our solar system.  The stiffness of spacetime is apparent if we consider a gravitational wave just barely detectable with current detectors, which periodically changes the difference in the distances along the arms of the detector by at most $10^{-19}$~m with a frequency of 100~Hz.  Such a signal has a flux (power per unit area) of about $10^{-5}~\text{W}/\text{m}^2$.  This is approximately the same flux in visible light measured 300 meters away from a standard 60 watt light bulb.  Thus even though these fluxes are equivalent, in the case of gravitational waves the signal is barely detectable with state of the art detectors, whereas with electromagnetic radiation the signal is easily detected by the human eye.

% paused here --dvs

\subsection{Data analysis}
  \label{subsec:DA}
The LVC detectors shown in Fig.~\ref{fig:IFOs} are measuring changes in the difference of the distances along the arms that are orders of magnitude smaller than the diameter of a proton (approximately $10^{-15}$~m), reaching sensitivities of $10^{-18}$ to $10^{-19}$~m.\cite{LIGO}  In addition to gravitational waves, such minute fluctuations in arm length can also be caused by many uninteresting sources, including seismic vibrations, local highway traffic, etc.  With so many noise sources causing signals at comparable levels to those we are trying to detect, advanced data analysis techniques are necessary. Many of these techniques rely on having a reasonably accurate theoretical model for the system emitting the gravitational waves, which provide the hypothesized signal that is then looked for in the data.  As described in Section~\ref{subsec:sources}, the LVC network is particularly sensitive to the mergers of black holes and neutron stars.  The main search algorithm for merger signals in the LVC uses the technique of matched filtering,\cite{S5lowMass} in which we first construct a bank of possible signals, and then search the data for instances of a statistically significant match to a signal in the bank.  This method is very efficient at detecting possible signals in large amounts of data, but does a poor job of determining source properties of individual signals, such as the masses of the merging objects and where in the sky the source is located.  To accurately estimate these parameters, other algorithms are required that are designed to analyze individual signals found by the main search algorithm.  These codes are based on Bayes' theorem,\cite{Bayes} and extract the maximum amount of information possible from the measured signal in the data, assuming the models used in the search accurately represent the signal in the data.\cite{MCMC1,MCMC2,nestedSamp}

\section{Gravitational-wave Science in the High School Classroom}
  \label{sec:classroom}
Over the course of the 2010--2011 academic year the authors worked to incorporate concepts and ideas from the field of gravitational-wave science into a high school astronomy curriculum.  Curriculum changes and lessons were implemented across 8 classes of students, with each class having about 25 students.  Each individual class had a mix of juniors and seniors, as well as non-honors and honors students.  Throughout the year new lesson topics were added pertaining to gravitational-wave astronomy.  Existing lessons were also modified and tied into gravitational-wave science (e.g., waves and interference), creating a common theme for the year's lessons.

In Section~\ref{subsec:waves} we describe how we incorporated gravitational-wave science into the existing unit on waves. In Section~\ref{subsec:GR} we present how we used gravitational waves as an introduction to the basics of general relativity. We also describe the demonstrations, guided inquiry, and manipulation of computational models that we designed and used in the classroom to support student learning of gravitational-wave science. In Section~\ref{subsec:signalProc} we discuss how we taught signal processing using Fourier techniques in the context of gravitational-wave science. Finally, in Section~\ref{subsec:otherTools} we provide a list of other tools for educators interested in teaching gravitational-wave science and in Section~\ref{subsec:future} we discuss future work.

\subsection{Waves and interference}
  \label{subsec:waves}
The physics of waves is relevant to both the gravitational waves themselves, and the design of the interferometric observatories built to detect them.  Many properties of electromagnetic waves, such as amplitude, frequency, and polarization, are also relevant to gravitational waves.  Since the Michelson interferometer is designed to utilize the properties of wave interference, a tabletop interferometer is an ideal demonstration to supplement discussion of these topics (see Fig.~\ref{fig:michelson}).  In this modified lesson on wave physics, we first reviewed transverse waves and their properties in a mini-lecture, material that the students had read the night before. After the review, students broke into small groups to work on an activity,\cite{waveActivity} making use of an online applet\cite{superpositionApp} that allowed students to manipulate overlapping waves and observe their resulting superposition.  After completing the activity, students were shown a tabletop Michelson interferometer, and were introduced to how constructive and destructive interference pertain to its design.  By applying very slight pressure to a mirror of the device, students developed an appreciation for the extreme sensitivity of the instrument.  Small changes in the interference pattern on the projection screen were observed even when students lightly tapped on the table holding the device.  On this particular apparatus there was a knob that would move the mirror by very small amounts. This knob was rotated back and forth as an exaggerated example of the effects of a gravitational wave on the interference pattern, while emphasizing to the students that during a gravitational-wave event, it is the distance between the mirrors and splitter that is expanding and contracting causing such an effect, not the physical movement of the mirrors. This mixture of hands-on demonstrations and computational models provided the students with a variety of ways to develop an understanding of wave physics.

Though we did not do so in this particular demonstration, future lessons could be improved by embedding a photodiode in the screen, which would measure fluctuations in the interference pattern.  By feeding to a speaker the varying voltage output by the photodiode, the changes in the interference pattern can be heard.  In particular this would make high frequency periodic changes to the pattern much easier to observe.  With this configuration we can potentially show that these instruments are even capable of picking up vibrations caused by speech, acting much like a microphone.

\subsection{General relativity and gravitational waves}
  \label{subsec:GR}
To help students begin to grasp the concept of gravity in the framework of general relativity, a large fabric sheet was used as a 3D representation of our 4D spacetime.  Spheres of various masses were placed on the sheet to demonstrate how the presence of mass curves spacetime. Rolling marbles near these massive objects then showed how this curvature acts in a way analogous to gravity. Before demonstrating each different scenario to the students, we asked for their predictions---what did they think would happen? In the first scenario, the sheet was empty and we gave a marble an initial velocity so as to send it in a straight path across the sheet. In the second scenario, a massive ball was placed in the middle of the sheet and the marble was given the same initial velocity as in the first scenario. Since the physics of objects on the fabric sheet is fairly intuitive to the students, they were able to predict that the marble's motion would be deflected by the curvature caused by the massive ball on the sheet. The students were also able to predict that the marble's trajectory would depend on its speed and distance of closest approach to the massive ball. To demonstrate circular motion of the marble around the massive ball (simulating, for example, the motion of planets around our sun), we found that it worked best to place the marble quite close to the massive ball and give it a small initial transverse velocity. The final demonstration was the case of a binary orbit, with two massive balls orbiting one another. During this demonstration students were asked to look at the behavior of the sheet far away from the binary system, so that they would observe the distant vibrations caused by the binary orbit. These vibrations, the students were told, were analogous to gravitational waves.

Following this discussion, students worked in small groups on an activity designed to guide them through the scientific reasoning required to determine the basic characteristics of a gravitational wave from a binary system.  This activity, which has been made publicly available,\cite{introGWactivity} begins by having the students analyze a simple sinusoidal wave, measuring its amplitude, period, and frequency.  They were then asked to draw examples of waves whose amplitudes and frequencies changed with time, to get them thinking about the possibility of waves without a fixed amplitude and frequency.  Students were then reminded that gravitational waves carry energy, and thus a system emitting gravitational waves must be losing energy.  This situation is analogous to what they had just seen in the fabric sheet demonstration, where instead of losing energy through gravitational waves the orbit lost energy due to friction.  This loss of energy resulted in the orbital radius shrinking and the orbital velocity increasing, just as it does during binary evolution.

The last part of the activity had students hypothesize what they believed a gravitational wave from an inspiraling binary system should look like, then compare that to a more rigorous computational model.\cite{pNapp} The applet used to interface with the computational model (provided for free by Wolfram Demonstrations\cite{wolfDemos}) displays the gravitational wave for a given set of parameters describing the binary system.  Slider bars then allow students to manipulate various parameters in the model (e.g., inclination, component masses, etc.), to investigate how these parameters can affect the modeled waveform.

This combination of visual demonstrations, guided inquiry, comparison and manipulation of computational models, and group discussion throughout is essential for having the students develop an accurate understanding of gravitational waves and overcome common misconceptions. We found that through the demonstration and the guided inquiry worksheet, over 80\% of students were able to predict and explain why the frequency of the gravitational wave will increase with time as the binary system inspirals. At this stage, however, only about half of the students were able to deduce that the amplitude of the wave will also increase with time, producing qualitatively accurate models like the one shown in Fig.~\ref{subfig:accurate}. The most common misconception was that the amplitude of the wave would remain constant with time, as illustrated in Fig.~\ref{subfig:misconception}. Through manipulating the computational model, students with the misconception recognized that the amplitude of the wave does in fact increase with time. However, even after this stage, many could not explain the physical cause for this amplitude increase. Only through a final full class discussion, in which students with more accurate models explained in their own words why the amplitude increases during the inspiral phase, did the remainder of the students understand this aspect of the system. More specifically, the students explained to their peers that the curvature of spacetime becomes more extreme as the massive objects move closer together (as seen in the fabric demonstration). As a result, we expect the strength of the gravitational wave (i.e., its amplitude) to also increase as the black holes approach one another (as seen in the computational model). The more accurate model thus shows both an increase in frequency and an increase in amplitude of the gravitational wave with time during the inspiral phase. This final discussion as well as the small group discussions were very lively. The students were clearly engaged and excited to explore and gain understanding of this cutting-edge science.  Additional lessons can later be used to show how such models are crucial to gravitational-wave detection in noisy data.\cite{icebreaker}

\subsection{Signal processing}
  \label{subsec:signalProc}
Signal processing is a skill common to many fields in observational science, but is often not taught, or even mentioned, until students are in college.  By teaching about gravitational waves elsewhere in the curriculum, we were able to provide an authentic context in which to learn about signal processing.  Not only that, but we were also able to appeal to many students' interest in music and music editing while doing so.

With such high levels of noise, signal processing and data analysis for gravitational-wave detectors is a challenging task.  One crucial tool in the search for any periodic signal in time-domain data is Fourier analysis.\cite{fourier}  Rather than scanning the data for indicators of the detector arm length changing periodically as a function of time (something made impossible by the noise causing such similar periodic changes), Fourier analysis allows one to transform the data to look at power as a function of frequency instead.  When viewing data in the frequency domain, the presence of a low-level periodic signal in random noise is easily seen, even with signal amplitudes much smaller than that of the noise.  In the context of gravitational waves, a given waveform is generated according to a computational model.  This model is then subtracted from the data, and the remaining data is compared to the noise measured in earlier data that did not contain a measurable signal.  How well these data with the model signal subtracted match the previously measured noise gives a quantitative way to assess how well the model matches the signal.

\begin{figure}[ht]
    \subfigure[~Time Domain Signal]{
   \includegraphics[width=3.05 in]{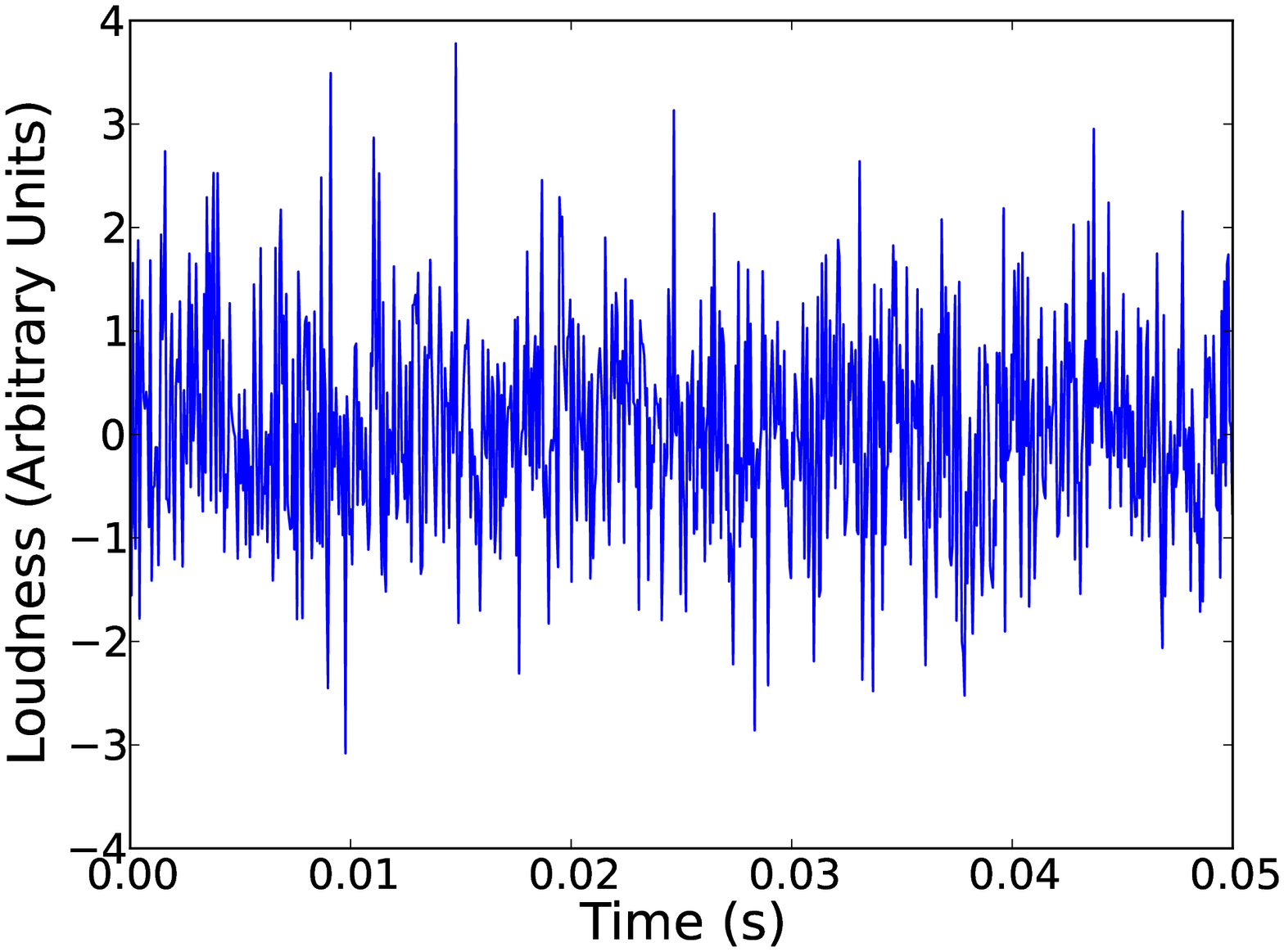}
    \label{subfig:audacityTD}
    }
    \subfigure[~Frequency Domain Signal]{
   \includegraphics[width=3.0 in]{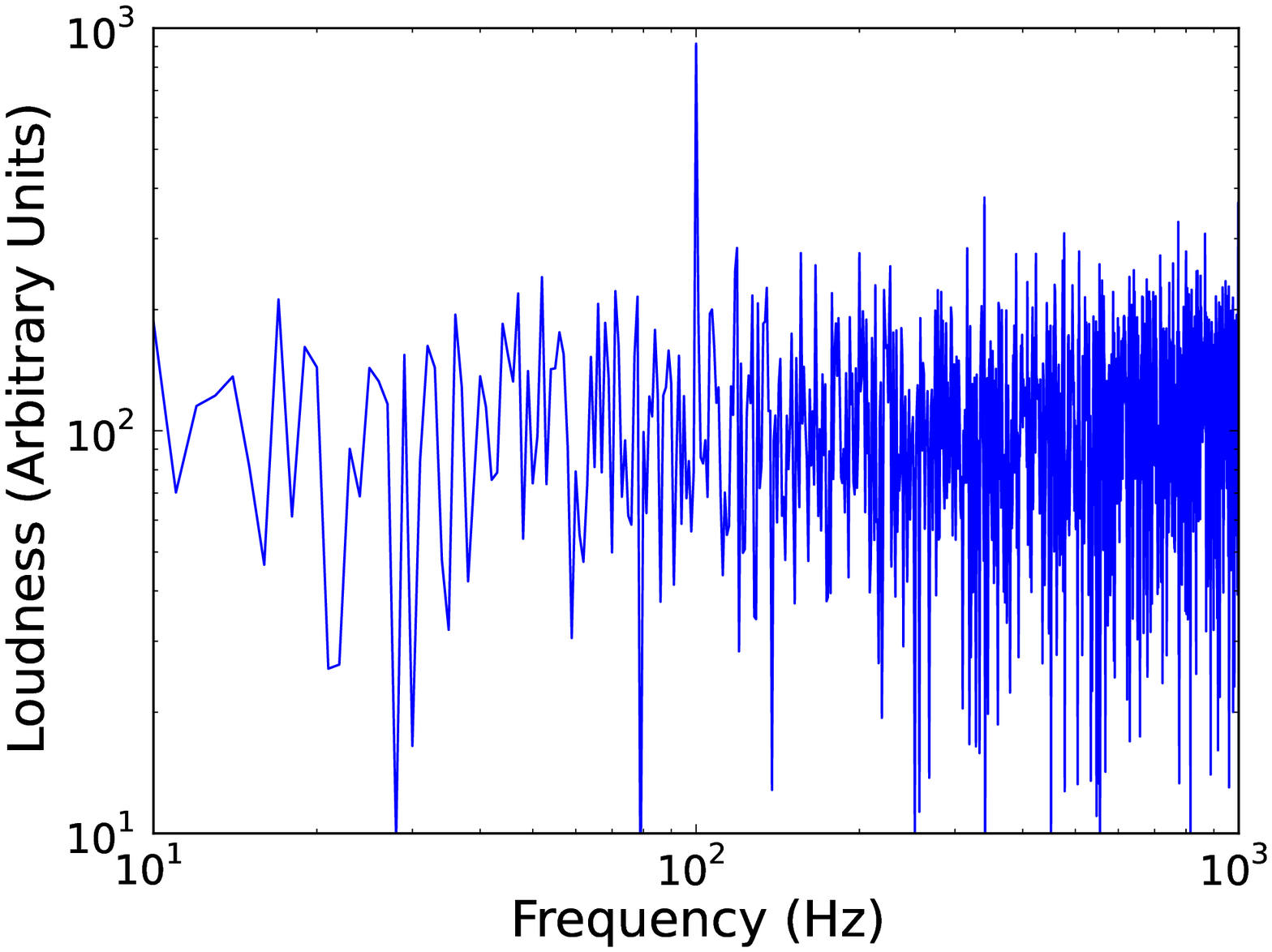}
    \label{subfig:audacityFD}
    }
    \caption{\label{fig:audacityScreenshots}Students used the open-source audio editing program Audacity to explore the 
              use of Fourier analysis in signal processing.  \subref{subfig:audacityTD} A composite 
              track containing white noise and a 100 Hz sine wave, generated by the student. 
              \subref{subfig:audacityFD} The result of a Fourier analysis of the combined signal and noise, showing
              the presence of a strong signal at 100 Hz.}
\end{figure}

Students used Audacity,\cite{audacity} an open-source audio editing program, to generate specific tone and noise spectra. They then added the tone and the noise spectra together. Figure~\ref{subfig:audacityTD} shows the time-domain track of the combined data, containing white noise and a low-amplitude 100 Hz sinusoidal wave. By using Audacity, students were able to see that a visual inspection of the time-domain data does not allow them to separate the signal from the noise, nor can they hear the signal when playing the audio sample. The students then took the Fourier transform of the composite track, as shown in Fig.~\ref{subfig:audacityFD}. The spike corresponds to the frequency of the input tone's low-amplitude sine wave. This demonstrated to them how using Fourier analysis can aid in the search for a quiet (low amplitude) coherent signal buried in random noise. This is a problem very similar to that encountered in gravitational-wave data analysis.

\begin{figure}[ht]
    \subfigure[~Simulated LIGO Noise Spectrum]{
   \includegraphics[width=3.0 in]{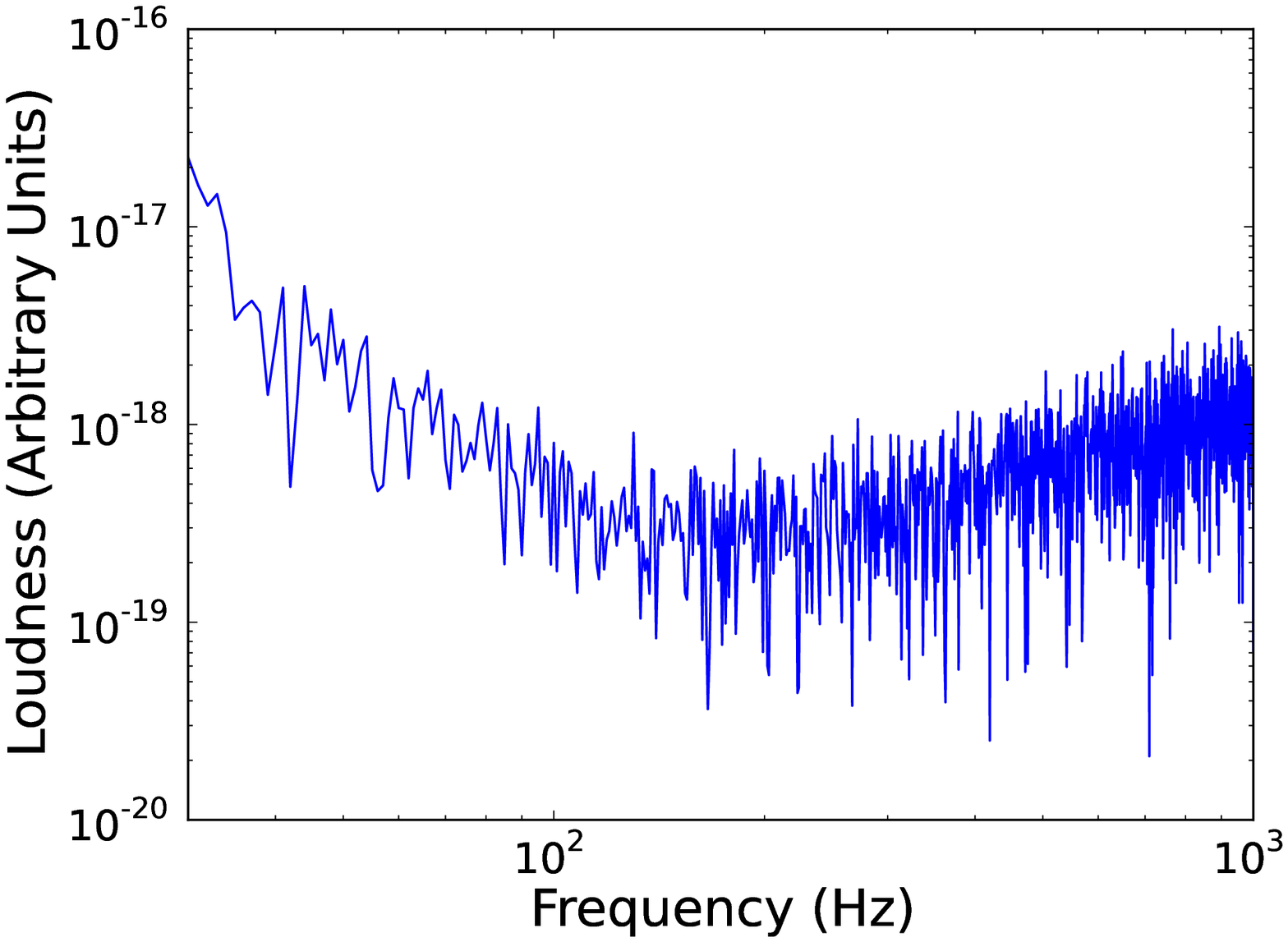}
    \label{subfig:noise}
    }
    \subfigure[~Spectrum of a Gravitational-wave Model]{
   \includegraphics[width=3.0 in]{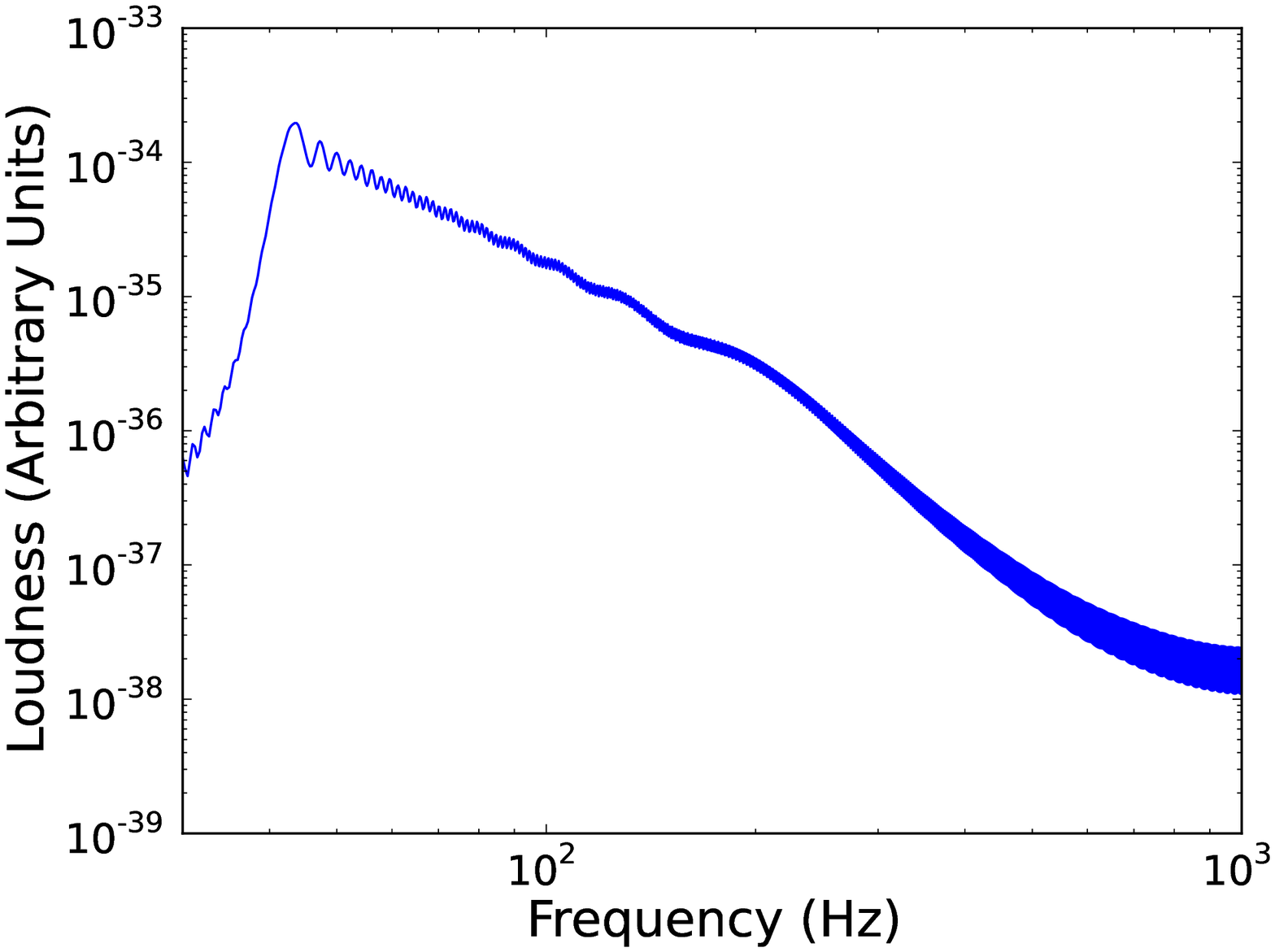}
    \label{subfig:signal}
    }
    \caption{\label{fig:LIGOsignals}Since LIGO is most sensitive to frequencies around 100 Hz, noise and signals relevant to LIGO
              can be converted to sound data that is audible to the human ear.  Using sound editing software,
              students were able to listen to and plot the spectra of \subref{subfig:noise} simulated LIGO noise,
              and \subref{subfig:signal} a gravitational-wave signal from a binary merger.}
\end{figure}

Finally, students were shown a simulated noise spectrum of LIGO, seen in Fig.~\ref{subfig:noise}, and the frequency-domain representation of a gravitational wave from a binary inspiral, shown in Fig.~\ref{subfig:signal}.  Students were also able to listen to the audio representation of these data, and could distinctively hear the difference between white noise and LIGO's noise, as well as listen to the waves they had learned about in a previous activity (see Section~\ref{subsec:GR}).

\subsection{Other tools for gravitational-wave education}
  \label{subsec:otherTools}
As part of its education and public outreach efforts, the LVC has created many other tools to assist with educating students about gravitational waves.  The LIGO Science Education Center\cite{LIGOscience} located on the Livingston, LA detector site hosts field trips and professional development workshops, as well as a Research Experience for Teachers program.\cite{RET}  This program is a six-week paid internship for K-12 teachers, designed to provide teachers with the opportunity to work in a scientific research environment on topics related to LIGO.  The Einstein's Messengers website\cite{einsteinMessanger} provides many curricular resources including connections to state and national standards, teacher and student study guides, and classroom activities related to LIGO and gravitational waves. Lastly there are many web applets and games designed to introduce students to concepts in gravitational-wave physics and detector technology.  The gwoptics.org website\cite{gwoptics} hosts outreach material, including several games developed by the gravitational-wave group in Birmingham, UK.  Black Hole Hunter\cite{blackholehunter} is a game intended to introduce people to gravitational-wave data analysis, having them search for the sound of gravitational waves in simulated noisy data.

\subsection{Future work}
  \label{subsec:future}
Now that basic gravitational-wave science has been integrated into the astronomy curriculum, more advanced topics can be introduced in coming years.  For example, the addition of independent study projects would provide another way to engage student in hands-on activities.  One possible project would be to have a group of students design and build their own interferometric detector.  The components necessary to construct a tabletop interferometer are relatively cheap, requiring only a laser, beamsplitter, photodiode, tabletop, and two mirrors.  An optical table would be ideal to house the interferometer, though that could be quite expensive to obtain.  Cheaper tables could be constructed from wood or other materials, and table construction itself could even be part of the project.  Once the detector is built, students could experiment with ways of making it more sensitive and reducing the effects of environmental noise.

The material presented here was integrated into the astronomy curriculum throughout the year.  Time should now be spent to condense this into a single self-contained unit that can be more easily inserted into a K-12 or introductory college physics curriculum.

One main area still to be added to the unit is parameter estimation.  After the detection of the first gravitational wave, the era of gravitational-wave astronomy will begin.  In order to extract all the information available from a gravitational-wave signal, one must use a technique specifically designed to do so.  Some work has been done to bring parameter estimation concepts into the classroom,\cite{PElesson} however lessons focusing on the currently used Bayesian methods have yet to be developed.  The Bayesian analysis techniques employed by the LIGO-Virgo collaboration are designed to extract the physical characteristics of the source of a measured gravitational wave.\cite{MCMC1,MCMC2,nestedSamp}  These methods utilize concepts from computational science, statistics, applied math, and other disciplines.  By including this in the curriculum, students would be exposed to topics across the STEM fields that they would normally never see in the high school setting, further demonstrating the importance of interdisciplinarity in science.

\section{Summary}
Gravitational-wave astronomers are on the verge of directly detecting gravitational waves for the first time.  Within the next decade, gravitational-wave science will change the field of astronomy by opening a new window to the universe.  As gravitational-wave science assumes a more prominent role in the astronomical community, it will be important for people to have at least a basic understanding of what gravitational waves are.  This article has provided several examples of how to introduce students to waves, interference, gravity, general relativity, and basic data analysis techniques through their applications to gravitational-wave science.  Through these interactive demonstrations and computationally based activities, students are able to learn about gravitational waves and the technology behind their detection, without losing focus on topics already in the curriculum.

\begin{acknowledgments}
This work was supported by the NSF GK-12 grant, award DGE-0948017, and the NSF LIGO grant, award PHY-0969820. The authors would like to thank the referees for their comments and suggestions.
\end{acknowledgments}

\theendnotes

\newpage
\end{document}